%% file: ms.tex
\documentclass[12pt,preprint2]{aastex}

\newcommand{\beq}{\begin{equation}}
  \newcommand{\eeq}{\end{equation}}
\newcommand{\bdm}{\begin{displaymath}}
  \newcommand{\edm}{\end{displaymath}}

\newcommand{\chan} {{\it Chandra}}
\newcommand{\xmm}  {{\it XMM-Newton}}

\newcommand{\cxo} {CXO}

\newcommand{\emd} {\mbox{$EMD$}}
\newcommand{\heg}  {HEG}
\newcommand{\rgs} {RGS}
\newcommand{\hetgs} {HETGS}

\newcommand{\meg}  {MEG}

\newcounter{ion}
\newcommand{\eli}[2]{\setcounter{ion}{#2}#1{~\sc\roman{ion}}}

\newcommand{\mone}{^{-1}}
\newcommand{\mtwo}{^{-2}}
\newcommand{\mthree}{^{-3}}

\begin{document}

\title{
  Evidence for Accretion in the High-resolution X-ray Spectrum of
  the T Tauri Star System Hen~3-600}

\author{
  David P. Huenemoerder\altaffilmark{1}, 
  Joel H. Kastner\altaffilmark{2}, 
  Paola Testa\altaffilmark{1}, 
  Norbert S. Schulz\altaffilmark{1}, 
  David A. Weintraub\altaffilmark{3}
  {\newline\smallskip{\small
  (Accepted for publication in the
  Astrophysical Journal, December 20, 2007, v671.)}}
}

\altaffiltext{1}{Massachusetts Institute of Technology,
  Kavli Institute for Astrophysics and Space 
  Research, 70 Vassar 
  St., Cambridge, MA, 02139}   
\altaffiltext{2}{Chester F. Carlson Center for
  Imaging Science, Rochester Institute of Technology,
  Rochester, NY 14623; jhk@cis.rit.edu}
\altaffiltext{3}{Dept.\ of Physics and Astronomy, Vanderbilt University,
  Nashville, TN, 37235}

\begin{abstract}
  
  We present high-resolution X-ray spectra of the nearby, multiple T
  Tauri star (TTS) system Hen~3-600, obtained with the High Energy
  Transmission Grating Spectrograph (\hetgs) aboard the \chan\ X-ray
  Observatory (\cxo).  Both principle binary components of Hen~3-600
  ($A$ and $B$, separation $1.4''$) were detected in the zeroth-order
  \cxo/\hetgs\ X-ray image.  Hen~3-600-$A$ --- the component with a
  large mid-infrared excess --- is a factor $\sim2-3$ fainter in
  X-rays than Hen~3-600-$B$, due to a large flare at Hen~3-600-$B$
  during our observation. The dispersed X-ray spectra of the two
  primary components overlap significantly, so spectral analysis was
  performed primarily on the first-order spectrum of the combined
  $(A+B)$ system, with analysis of the individual dispersed spectra
  limited to regions where the contributions of $A$ and $B$ can be
  disentangled via cross-dispersion profile fitting. This analysis
  results in two lines of evidence indicating that the X-ray emission
  from Hen~3-600 $A+B$ is derived, in part, from accretion processes:
  (1) The line ratios of He-like \eli{O}{7} in the spectrum of
  Hen~3-600 $A+B$ indicate that the characteristic density of its
  X-ray-emitting plasma is significantly larger than those of
  coronally active main sequence and pre-main sequence stars. (2) A
  significant component of low-temperature (2--3 MK) plasma is present
  in the Hen~3-600 $A+B$ spectrum; this ``soft excess'' appears
  somewhat stronger in component $A$. These results for Hen~3-600
  $A+B$ are consistent with, though less pronounced than, results
  obtained from X-ray gratings spectroscopy of more rapidly accreting
  TTS systems.  Indeed, all of the emission signatures of Hen~3-600
  $A+B$ that are potential diagnostics of accretion activity --- from
  its high-resolution X-ray spectrum, through its UV excess and
  H$\alpha$ emission line strengths, to its weak near-infrared excess
  --- suggest that its components (and component $A$ in particular)
  represent a transition phase between rapidly accreting, classical T
  Tauri stars and non-accreting, weak-lined T Tauri stars.

\end{abstract}


\section{Introduction}\label{sec:intro}

X-ray observations of star formation regions have established that
low-mass, pre-main sequence (T Tauri) stars are luminous X-ray sources
\citep{Feigelson:Montmerle:1999}. Such high-energy radiation may exert
a profound influence over the process of planet formation
\citep[e.g.][]{Glassgold:al:2004, Feigelson:Getman:al:2005}.  The
origin of pre-main sequence (pre-MS) X-ray emission is still the
subject of debate, however.  Although most recently formed, low-mass
stars likely emit X-rays as a consequence of solar-like coronal
activity
\citep{Kastner:Huenemoerder:al:2004,Preibisch:Kim:al:2005,Telleschi:Gudel:Briggs:Audard:Palla:2007}
we and others have suggested that the X-ray emission from certain
actively accreting pre-main sequence stars may be a direct result of
mass accretion from a circumstellar disk onto the forming star
\citep{Kastner:02,Kastner:Richmond:al:2004,Kastner:Richmond:al:2006,
  Stelzer:Schmitt:2004,Schmitt:Robrade:al:2005,Gunther:Liefke:al:2006,
  Argiroffi:Maggio:Peres:2007,Telleschi:Gudel:Briggs:Audard:Scelsi:2007}.

To improve our understanding of the physical conditions within X-ray
emitting regions of both actively accreting, classical T Tauri stars
(cTTS) and apparently non-accreting, weak-lined T Tauri stars (wTTS),
we are conducting a high-resolution X-ray spectroscopic study of the
TW~Hya Association \citep[TWA;][]{Kastner:Zuckerman:1997,
  Webb:Zuckerman:al:1999, Zuckerman:Webb:2001,
  Song:Zuckerman:Bessell:2004, Song:Zuckerman:Bessell:2003} with the
\chan\ X-ray Observatory's High Energy Transmission Grating
Spectrometer \citep[\hetgs;][]{HETG:2005}.  The TWA is uniquely well
suited to this study, due to its proximity (typical member distances
$D\sim50$ pc) and lack of intervening cloud material, as well as its
status as an evolutionary ``missing link'' between embedded pre-MS
stars and the zero-age main sequence
(\citet{Kastner:Crigger:al:2003,delaReza:al:2006}; also see the review
by \citet{Zuckerman:Song:2004}).

Observations of TWA members with \hetgs\ provide unique plasma
diagnostics that have generated insight into the X-ray emission
mechanisms of T Tauri stars. In particular, the density-sensitive line
ratios of He-like ions (\eli{O}{7}, \eli{Ne}{9}) detected in the
\chan/\hetgs\ X-ray spectrum of TW~Hya indicated plasma densities in
the range $11.9 < \log n_e < 12.5$
\citep{Kastner:02,Testa:Drake:al:2004b}. This is roughly an order of
magnitude larger than all active main sequence stars that have been
observed at high spectral resolution
\citep{Testa:Drake:al:2004b,Ness:gudel:al:2004}. Similar results for
density-sensitive line ratios subsequently have been obtained from
X-ray gratings spectroscopy of the cTTS BP Tau
\citep{Schmitt:Robrade:al:2005}, V4046 Sgr
\citep{Gunther:Liefke:al:2006}, and MP Mus
\citep{Argiroffi:Maggio:Peres:2007}.  Furthermore, the plasma
temperature measured from \hetgs\ and \xmm\ spectra of TW~Hya, $\sim3$
MK, is consistent with adiabatic shocks arising in gas at free-fall
velocities expected for accretion onto the star \citep{Kastner:02,
  Stelzer:Schmitt:2004}.

In stark contrast, the \hetgs\ spectrum of the wTTS system HD~98800, a
member of the TWA and which is apparently non-accreting, closely
resembles those of active main sequence stars
\citep{Kastner:Huenemoerder:al:2004}.  In particular, its He-like
\eli{Ne}{9} and \eli{O}{7} triplet line ratios and, hence, plasma
densities are well within the range characteristic of ``classical''
coronal sources, and its plasma temperature distribution is
significantly hotter than that of TW~Hya.

Typical signatures of active accretion are strong H$\alpha$ emission,
ultraviolet excess from an accretion disk boundary layer, and a
near-infrared excess from hot dust in the inner disk.  To fully
understand the dependence of high energy emission on accretion rate --
if any -- we need high resolution X-ray spectra from objects which
span a large range in the optical, UV, and IR diagnostics.  Here we
describe the results of \hetgs\ observations of TWA member Hen~3-600,
a multiple pre-MS star system whose principle components are separated
by $1.4''$ \citep{Webb:Zuckerman:al:1999,
  Jayawardhana:Hartmann:al:1999}.  Like HD~98800, one primary
component of Hen~3-600 evidently is surrounded by a dusty disk, based
on its IR spectral distribution \citep{Jayawardhana:Hartmann:al:1999}.
The mid-infrared spectrum of Hen~3-600, like that of TW~Hya, displays
evidence for crystalline silicate dust grains
\citep{Honda:Kataza:al:2003, Uchida:Calvet:al:2004}, suggesting
material in the circumstellar disk of Hen~3-600-$A$ has undergone
considerable processing and evolution.

The presence and strength of H$\alpha$ emission and UV excesses in the
spectra of TW~Hya and Hen~3-600-$A$ (Table~\ref{tbl:stars}) indicate
that accretion is ongoing in each case.  Indeed, apart from TW~Hya
itself, Hen~3-600-$A$ is the strongest H$\alpha$ emission source in
the TWA \citep{Webb:Zuckerman:al:1999}.  Based on H$\alpha$ line
profile analysis, both stars are accreting at rates significantly
lower than those of $\sim1$ Myr-old TTS that are still found in their
nascent dark clouds, with the accretion rate of Hen~3-600-$A$ about an
order of magnitude lower than that of TW~Hya
\citep{Muzerolle:Calvet:2000}.  This is consistent with the relatively
small UV excess of Hen~3-600, as UV excess is also widely used as an
indicator of TTS accretion rate.  TW~Hya is known to have a nearly
face-on disk \citep{Krist:Stapelfeldt:2000,Kastner:Zuckerman:1997} and
is thus ideal for probing accretion since there is no intervening
local X-ray absorption by neutral hydrogen, and because the
high-latitude accretion region (if directed by bipolar magnetic
fields) is in full view.  We have no direct measurement of the
inclination of the Hen~3-600 components to our line-of-sight, but we
infer that it is also low, given its combination of large mid-IR
excess and negligible optical reddening; its measured value of $(B-V)
= 1.52$ is nearly that of an unreddened M3 photosphere
\citep{Johnson:1966}.  The contrasts in accretion diagnostics apparent
in Table~\ref{tbl:stars}, and the status of Hen~3-600 as an
intermediate case between TW~Hya and HD~98800, make Hen~3-600 a key
TWA system for further study with \chan/\hetgs.  Both components, $A
\& B$, are considered members of the TWA based on photometric,
spectroscopic, and kinematic criteria \citep{Torres:Guenther:al:2003,
  Reid:2003, Webb:Zuckerman:al:1999}.

\section{Observations, Data Reduction, and Results}\label{sec:obs}

We observed Hen~3-600 with \chan/\hetgs\ for 101 ks starting on 2004 March
16 (observation identifier 4502) in the default configuration
(timed exposure, ACIS-S detector array) and under nominal operating
conditions.  Data were re-processed with \chan\ Interactive Analysis
of Observations software (CIAO, version 3) to apply updated
calibrations (CALDB 3.2). Events were also cleaned of the detector
artifacts on CCD 8 (``streaks'').  The effective exposure time after
data reprocessing is 99.4 ks. 
Spectral responses, including corrections
for ACIS contamination, were generated with CIAO.

The resulting \chan/\hetgs\ spectral image of the Hen~3-600-$A/B$ binary
yielded 883 and 1610 zeroth order counts from $A$ and $B$,
respectively, and 3420 combined (\meg\ and \heg) first order counts.
The dispersed spectra are not well resolved spatially over the entire
spectral range (see \S~\ref{sec:xdisp}). To estimate count rates for
the individual components of the Hen~3-600 binary, we extracted
separate dispersed spectra within cross-dispersion regions centered on
$A$ and $B$, with the corresponding zeroth-order centroids as the
wavelength reference.  We thus estimate combined \meg\ and \heg\ plus
and minus first order count rates of 0.010 s$^{-1}$ and 0.024 s$^{-1}$
for components $A$ and $B$, respectively; these estimates are subject
to some uncertainty, however, given the overlap between the dispersed
spectra of $A$ and $B$ in the cross-dispersion direction.

Line positions and fluxes were measured with
ISIS\footnote{\raggedright ISIS is available from {\tt
    http://space.mit.edu/CXC/ISIS}} \citep{Houck:00,Houck:2002} by
convolving Gaussian profiles with the instrumental response.
Determination of an accurate continuum is important. Even at \hetgs\
resolution, there are portions of the spectrum for which the local
minima are a poor approximation to the true continuum.  Hence, we
adopt an iterative procedure whereby we first use a low order
polynomial to fit the local continuum knowing it to be in error in
some spectral regions. After we obtain a plasma model from line
fluxes, we evaluate a global continuum and re-fit the lines and
continue to iterate in this fashion if required.


\subsection{Zeroth-order image, light curves, and CCD spectra}\label{sec:zodata}

\begin{figure*}[t]
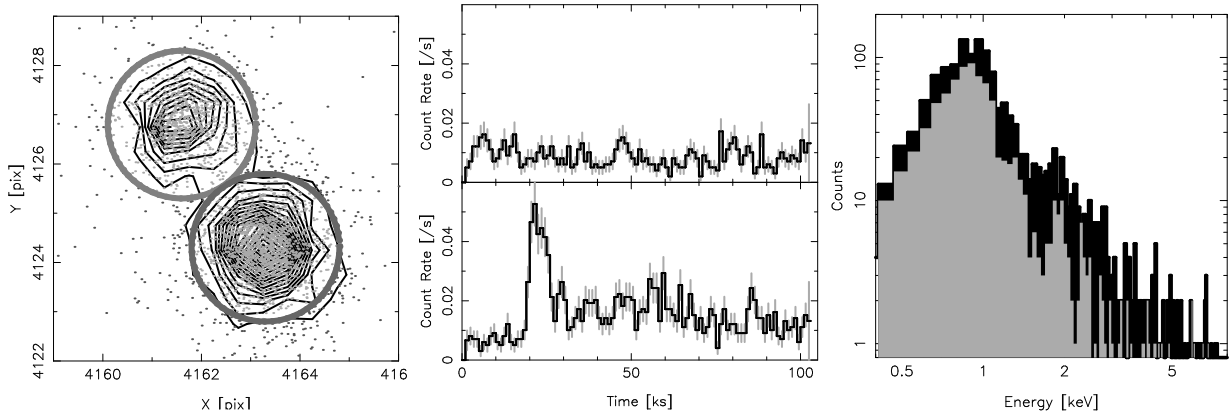

  \epsscale{0.7}
  \centerline{\plotone{f1a.eps}\plotone{f1b.eps}\plotone{f1c.eps}}
  \caption 
  { \small The zeroth order data for Hen~3-600. The left panel shows the
    central field in sky pixel coordinates as contours with individual
    events over-plotted as points.  The solid-outlined circles are the
    0.7'' (1.4 pixel) radius extraction regions for zero order light
    curves and spectra. The outer contour is for 1 count, and the
    interval is 10 counts, with the maximum counts in the brighter
    source being about 180.  The upper left source is component
    $A$.  The middle two-panel plot shows the zero order light curves
    for $A$ (upper) and $B$ (lower) in 1 ks bins, for the CCD {\tt
      ENERGY} range of 0.25--7.0 keV; backgrounds are
    negligible.  The right plot shows the zero order CCD spectra, with
    the brighter source $B$ shaded in black, and source $A$, which is
    fainter and somewhat softer, in gray.  CCD photon pileup was not
    significant at these low rates.
    \label{fig:zo}
  }
\end{figure*}

\begin{figure*}[t]
  \epsscale{2.33}
  \centerline{\plottwo{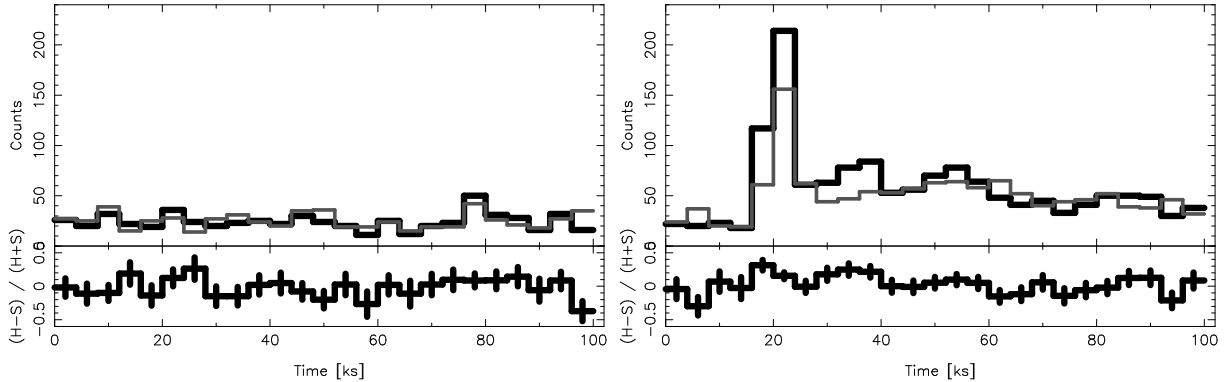}{f2b.eps}}
  \caption{\small Light  curves (top panels) and hardness ratios (bottom
    panels) for components $A$ (left) and $B$ (right) obtained from
    the dispersed $\meg+\heg$ spectral events for positive and
    negative orders 1--3.  In the top panels, the lighter (gray)
    histograms indicate the 1.7--10 \AA\ (``hard'') band, and the
    darker (black) histograms indicate the 10--26 \AA\ (``soft'')
    spectral region. The bottom panels display the time
    history of hardness
    ratios constructed from the counts in these ``soft'' and
    ``hard'' bands. Since the dispersed spectra of $A$ and $B$ are 
    not fully resolved, these curves were extracted from narrow
    regions centered on the component's spectrum; each is contaminated
    to some degree by the other source.
    \label{fig:tglc}
  }
\end{figure*}

In Figure~\ref{fig:zo} we display the zeroth-order \hetgs\ image,
light curves, and CCD energy spectra of the Hen~3-600 system.  Both
primary components were detected, with separation ($1.4''$) and
orientation in excellent agreement with earlier infrared imaging
\citep{Jayawardhana:Hartmann:al:1999}.  Hence, there is no ambiguity
in attributing the X-ray emission to components $A$ and $B$.  However,
the X-ray flux ratio of the $AB$ pair is the reverse of that in the
optical and infrared. In those wavelength regimes, component $A$ is
brighter --- indeed, its infrared excess likely dominates that of the
Hen~3-600 system \citep{Jayawardhana:Hartmann:al:1999}. During our
observations, however, $B$ was the brighter X-ray source.  

The light curve shown in Figure~\ref{fig:zo} reveals that this X-ray
mean flux difference in Hen~3-600 is due to a large flare
that occurred on component $B$, $\sim16$ ks after the start of the
observation.  This flare displayed a steep ($<2$ ks) rise and large
amplitude, achieving a count rate a factor $\sim5$ above pre-flare
rate.  The flare decays abruptly over $\sim10$ ks, but did not return
to the pre-flare level, but to one about two times higher.  

The count rate of component $A$ remained at a low level with small
fluctuations throughout our observation.  It was actually the brighter
of the two main components during the initial $\sim15$ ks of the
\hetgs\ observation.  A similar reversal in X-ray to optical and IR
occurred in HD~98800 in which the component surrounded by a dusty
disk, defined by presence of a strong IR excess, was also the weaker
and steadier X-ray source of the system
\citep{Kastner:Huenemoerder:al:2004}.

Component $B$ appears to be the harder of the two components in the
zeroth-order CCD spectra, which are also shown in Figure~\ref{fig:zo}.
In particular, the zeroth-order spectral energy distributions of $A$
and $B$ peak near $\sim0.9$ keV, but component $A$ drops off more
rapidly at higher energies.  Component $B$ has a weak but significant
Fe K$\alpha$ line at $\sim6.7$ keV.  

There is little evidence that the spectrum of $B$ hardened during the
flare.  The hardness increased only marginally from 15--40 ks from
observation start (Figure~\ref{fig:tglc}).  This behavior appears
distinct from that of coronal sources, which usually show much more
obvious hardening during flares, as in the case of HD~98800
\citep{Kastner:Huenemoerder:al:2004} and other coronally active stars
\citep[e.g.][]{Gudel:Audard:al:2004, Huenemoerder:Testa:al:2006}.  The
hardnesses of $A$ and $B$ outside the flare (excluding the 15--40 ks
interval) are similar.  This suggests that excluding the flare, that
the spectral energy distributions are similar. High resolution
spectral diagnostics (discussed later) reveal some differences between
$A$ and $B$.

We did not use the zero order for detailed spectral modeling, but
deferred to the high resolution spectrum.  We did evaluate the high
resolution-derived model for the sum of $A+B$ with the zero order
response to examine consistency with the sum of the zero order
spectra, but calibration systematic residuals make comparison
difficult. We thus only used the zero orders for variability and
qualitative assessment of the individual components.

\subsection{First Order Spectrum}\label{sec:tgspec}

\begin{figure*}[t]
  \centerline{\includegraphics[scale=1.0]{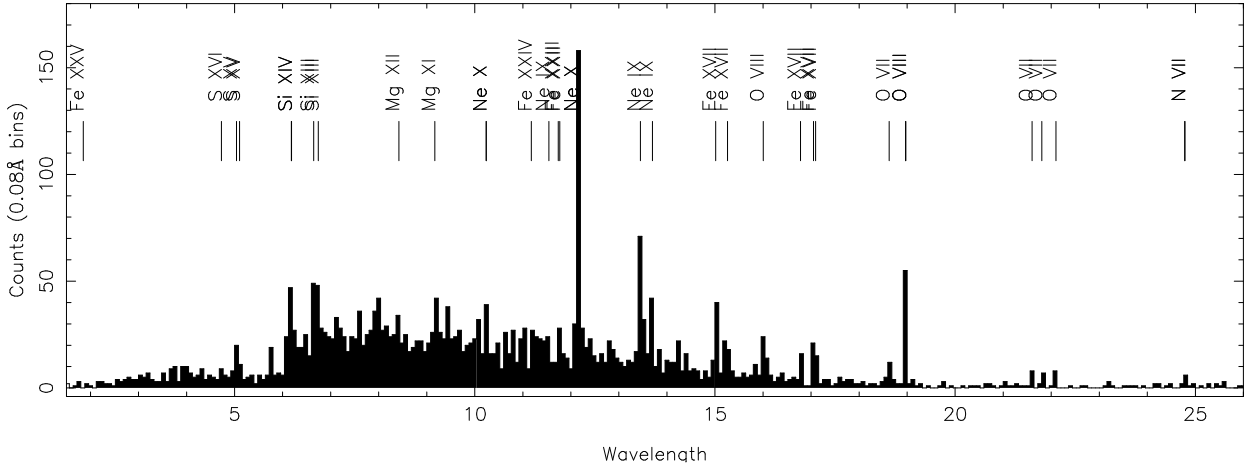}}
  \caption{\small The combined $\heg + \meg$ spectrum of $A+B$, heavily
    binned ($0.08$ \AA/bin).  Some prominent or important
    lines are marked.
    \label{fig:tgspec} }
\end{figure*}
%
The combined, first-order MEG and HEG spectrum of Hen~3-600 is
displayed in Figure~\ref{fig:tgspec}.  
As is also the case in HETGS spectra of fellow TWA members TW~Hya and
HD~98800 \citep{Kastner:02,Kastner:Huenemoerder:al:2004}, the most
prominent lines in the HETGS spectrum of Hen~3-600 are those of highly
ionized Ne and O, while lines of Fe are relatively weak.  We have
measured line fluxes by fitting the combined $A+B$ counts
(Table~\ref{tbl:lfluxemd}); for stronger lines we have determined the
relative contribution of each component (Table~\ref{tbl:abfluxes}).
In the following sections, we describe these determinations in more
detail.

\section{Spectral Analysis}

\subsection{Emission measure distribution}\label{sec:emd}

To ascertain the ranges of temperature and elemental abundances that
characterize the X-ray-emitting plasma of the Hen~3-600 system, we fit
its combined ($A+B$) dispersed HETGS spectrum with an emission measure
distribution ($EMD$) model.  The $EMD$ is a one-dimensional
characterization of a multi-thermal plasma.  It is a weighting
function which relates the line emissivities ($\epsilon$) to flux
($f$) via an integral equation,
$$ f_{jZ} = \frac{A_Z}{4\pi d^2} \int{ \epsilon_{jZ}(T)\, [ n_e n_H
  \frac{dV}{dT} ]\, dT }$$ \label{eq:emd}
for ion $j$ of element atomic number $Z$ and abundance
$A_Z$; $T$ is the plasma (electron) temperature, $V$ the
volume, $n_e$ and $n_H$ are the electron and hydrogen
densities, and $d$ is source distance.  The quantity in
square brackets is the $EMD$ (also commonly called the
differential emission measure [$DEM$]).  We have ignored any
density dependence on $\epsilon$, which in principle exists
and which would introduce another differential term into the
equation.  The few strongly density-sensitive lines of the
He-like triplets are handled independently. We also assume
uniform abundances, such that $A_Z$ appears outside the
integral.

We applied methods similar to those described in
\citet{Huenemoerder:Testa:al:2006} to perform a regularized fit of the
emission measure and abundances to line fluxes.  There are many
caveats to emission measure modeling \citep[see][for discussion and
further citations]{Huenemoerder:Testa:al:2006}.  Here we have the
added complication that components $A$ and $B$ are not fully resolved,
so we must fit their summed emission. The results for the $EMD$ of
Hen~3-600 hence implicitly rely on the assumption that the abundances
are homogeneous and identical in each component.  Lines used in the
$EMD$ fitting are listed in Table~\ref{tbl:lfluxemd}.  Those lines
with measured flux but which are known to be density sensitive
(He-like triplet intercombination and forbidden lines) or which had
large residuals (due to some systematic error, such as
misidentification) have their predicted flux enclosed in parentheses.

\begin{figure}[htb]
  \epsscale{0.9}
  \centerline{\plotone{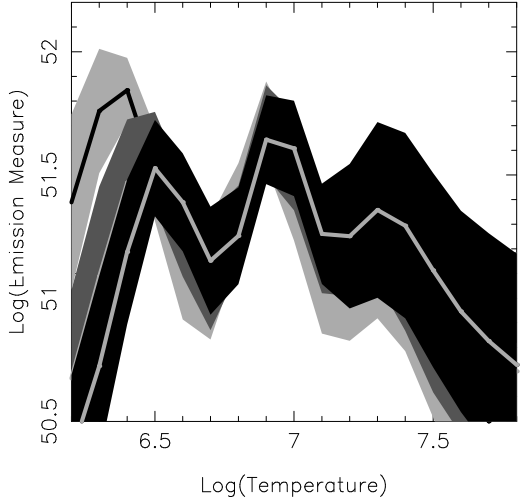}}
  \caption{\small
    Emission measure distribution derived from line  flux fitting to
    the $A+B$ dispersed spectrum.  The darkest shading  
    with the light  central line is the preferred solution for a
    hydrogen  column density of $N_H=10^{20}\,\mathrm{cm\mtwo}$, which 
    is effectively a negligible term for the line    uncertainties
    here.  The lightest shading seen at higher  emission 
    measure at low temperature (with dark central line) is for
    $N_H=3\times10^{21}$ cm$^{-2}$, which can 
    be excluded because of the implausibly large    abundance of
    nitrogen required to fit the $\lambda24.8$ line.    Intermediate
    shading is for $N_H=1\times10^{21}$    cm$^{-2}$.  
    \label{fig:emd}
  }
\end{figure}

Figure~\ref{fig:emd} shows the resulting $EMD$ for different assumed
values of the absorbing column ($N_H$).  The upper and lower bounds of
the shaded regions result from 100 Monte-Carlo iterations in which the
line fluxes were randomly perturbed within their measured
uncertainties.  Regardless of the assumed value of $N_H$, we find a
broad $T$ distribution with strong peaks at $\log T = 6.5$ and $\log T
= 7.0$, and a ``hot tail'' above $\log T = 7.4$.  The longest
wavelength (and characteristically cooler) lines are most sensitive to
the assumed $N_H$.  Because there are few counts in this
long-wavelength regime, we obtain effectively identical solutions for
values of $N_H \leq 10^{21}$ cm$^{-2}$.  If we assume a value of $N_H
= 3\times10^{21}$ cm$^{-2}$, then the $EMD$ rises strongly below $\log
T=6.5$; however, in this case the required N abundance is
unrealistically large, at 4.7 times Solar.  Hence, we adopt a value of
$N_H = 10^{20}$ cm$^{-2}$.  This value has $<10\%$ effect on line
fluxes in the \hetgs\ bandpass, and is consistent with the apparent
lack of interstellar reddening toward the TWA
\citep{Kastner:Crigger:al:2003} and with the apparently negligible
$E(B-V)$ toward and Hen~3-600 in particular.  The small inferred value
of $N_H$ (see also \S\ref{sec:ovii}) furthermore suggests we observe
the dust-disk-enshrouded component, $A$
\citep{Jayawardhana:Hartmann:al:1999}, at low inclination

\begin{figure}[t]
  \epsscale{0.9}
  \centerline{\plotone{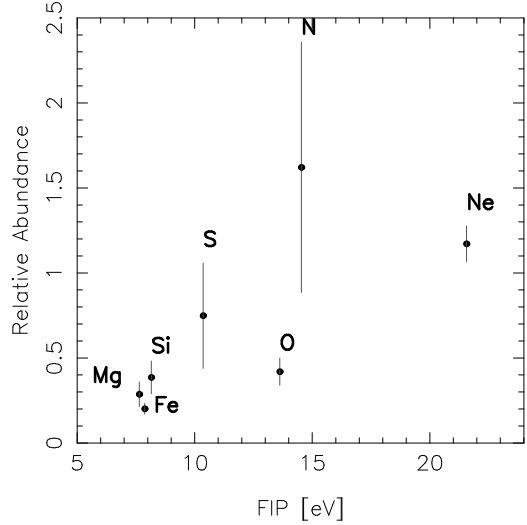}}
  \caption {\small Elemental abundances (relative to Solar) derived
    from the line-based emission measure fit, plotted against
    the First Ionization Potential (FIP).
    \label{fig:abundemd}
  }
\end{figure}

\begin{figure*}[h]
  \centerline{\includegraphics[scale=0.9]{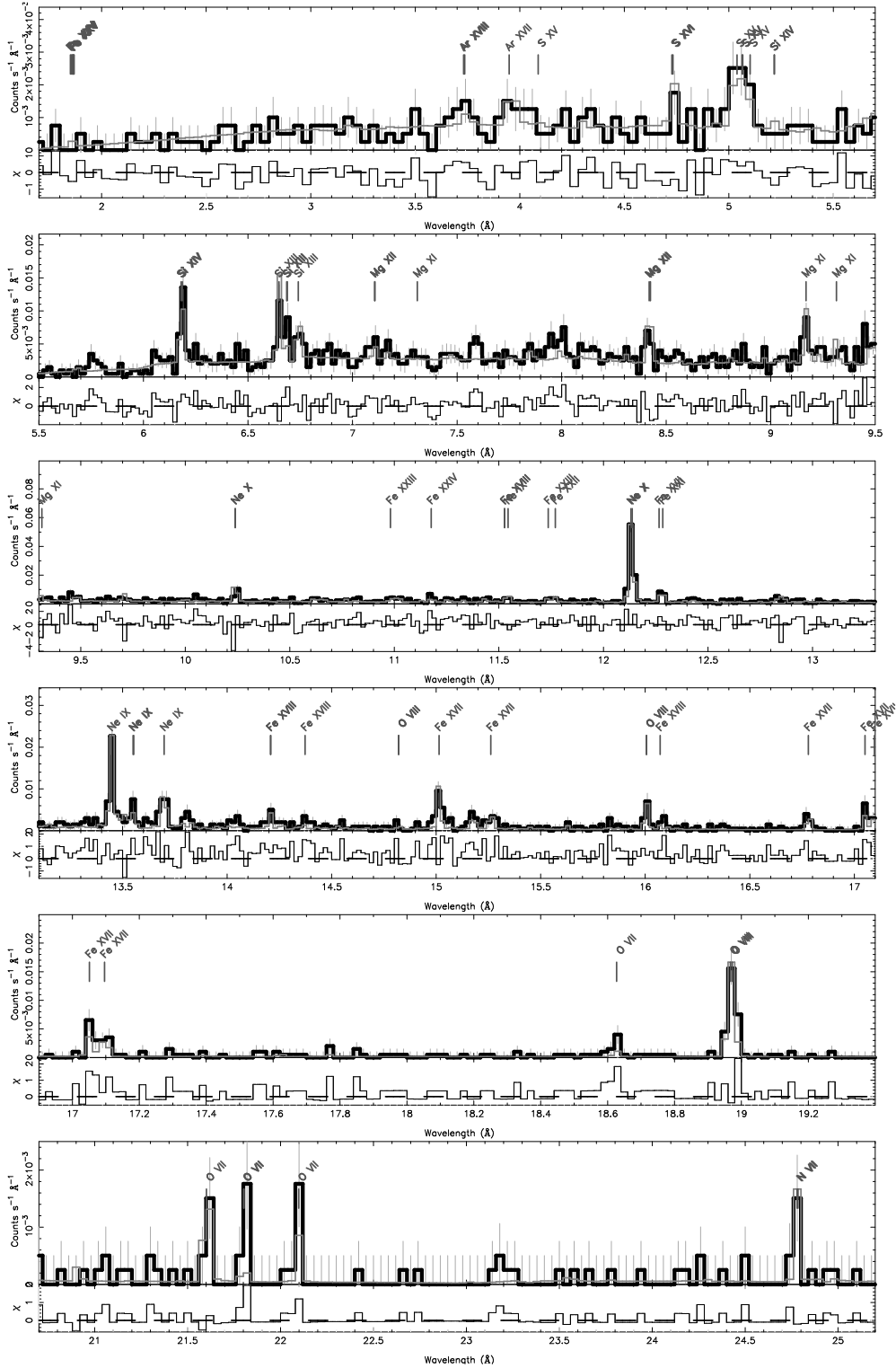}}
  \caption{\small The combined $\heg + \meg$ spectrum of $A+B$ (dark
    histogram), the model (light histogram), and $\delta\chi$
    residuals (lower panels).  Lines measured for emission measure
    analysis are labeled, and line details can be found in
    Table~\ref{tbl:lfluxemd}.  
    \label{fig:emdspec} }
\end{figure*}

The abundances obtained from the $EMD$ fitting under the assumption
$N_H=10^{20}$ cm$^{-2}$ are listed in Table~\ref{tbl:abundemd} and
displayed in Figure~\ref{fig:abundemd}.  We have adopted the
abundances of \citet{Anders:89} for our ratios to Solar values.
Figure~\ref{fig:emdspec} shows the observed spectrum overlaid with the
synthetic spectrum predicted by the emission measure distribution and
abundance model.

Assuming $N_H = 10^{20}\, \mathrm{cm^{-2}}$, the integrated flux
(1.5-25 \AA) obtained from the $EMD$-reconstructed model is
$1.2\times10^{-12}\,\mathrm{ergs\,cm^{-2}\,s^{-1}}$
($0.74\times10^{-3}$ $\mathrm{phot\,cm^{-2}\,s^{-1}}$), corresponding
to a luminosity $L_x=3.1\times10^{29}\,\mathrm{ergs\,s^{-1}}$ at the
estimated distance to Hen~3-600 of 45 pc, the mean of the values
derived by \citet{Kastner:Zuckerman:1997} ($39\pm 7$ pc, a
``photometric'' distance) and \citet{Webb:Zuckerman:al:1999} (50 pc,
an assumed value, based on Hipparcos distances of 47--67 pc measured
for other TWA members).
%

\clearpage
\clearpage

\subsection{Cross-dispersion decomposition}\label{sec:xdisp}

\begin{figure}[h]
  \centerline{\includegraphics[scale=0.8,angle=0]{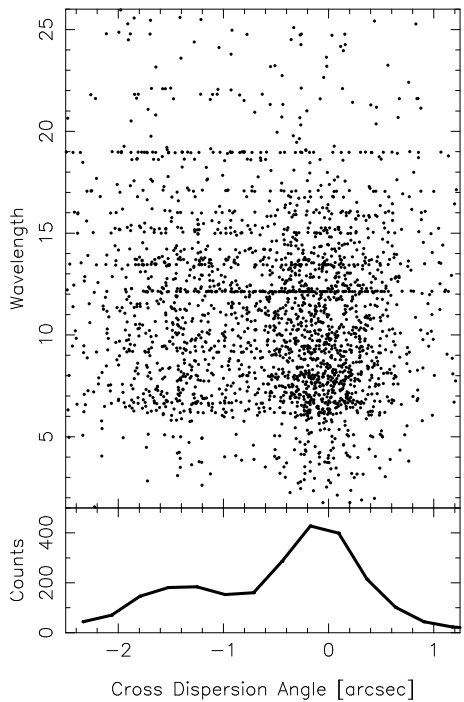}}
  \caption{\small The cross-dispersion distribution of \meg\ first order
    events.  The top panel shows a scatter plot of the events, and the
    bottom shows their cumulative histogram.  The grating
    cross-dispersion angle has been scaled to the imaging focal
    length.  In detail, this distribution depends upon wavelength.  On
    the average, for an extraction region boundary at about $-1$
    arcsec, about $7\%$ of the counts on the $A$-side (left) come from
    $B$, and there is $\sim4\%$ contamination in $B$ from $A$.
    \label{fig:rddist} }
\end{figure}
%
Though the main components of the Hen~3-600 binary system are well
resolved in the zeroth-order image (see Figure~\ref{fig:zo}),
the dispersed spectra of the two components are only marginally
resolved spatially in the cross-dispersion direction
(Figure~\ref{fig:rddist}).  The degree of cross-contamination of the
spectrum increases with wavelength due to the astigmatism inherent in
a Rowland spectrometer.  To obtain emission line fluxes and ratios in
a consistent manner across the entire spectrum we have modeled the
cross-dispersion profiles in the first-order spectra as a function of
wavelength to quantify the relative contribution of each source to 
the brighter emission lines.

Templates for cross-dispersion profiles were based on the same MARX
simulations from which the \chan\ calibration database line spread
fuctions were derived \citep{Marshall:Dewey:Ishibashi:2004}.  The
cross-dispersion profile of a single source is characterized by the
sum of two Gaussian plus two Lorentzian distributions whose parameters
depend on the grating type and wavelength.

\begin{figure*}[t]
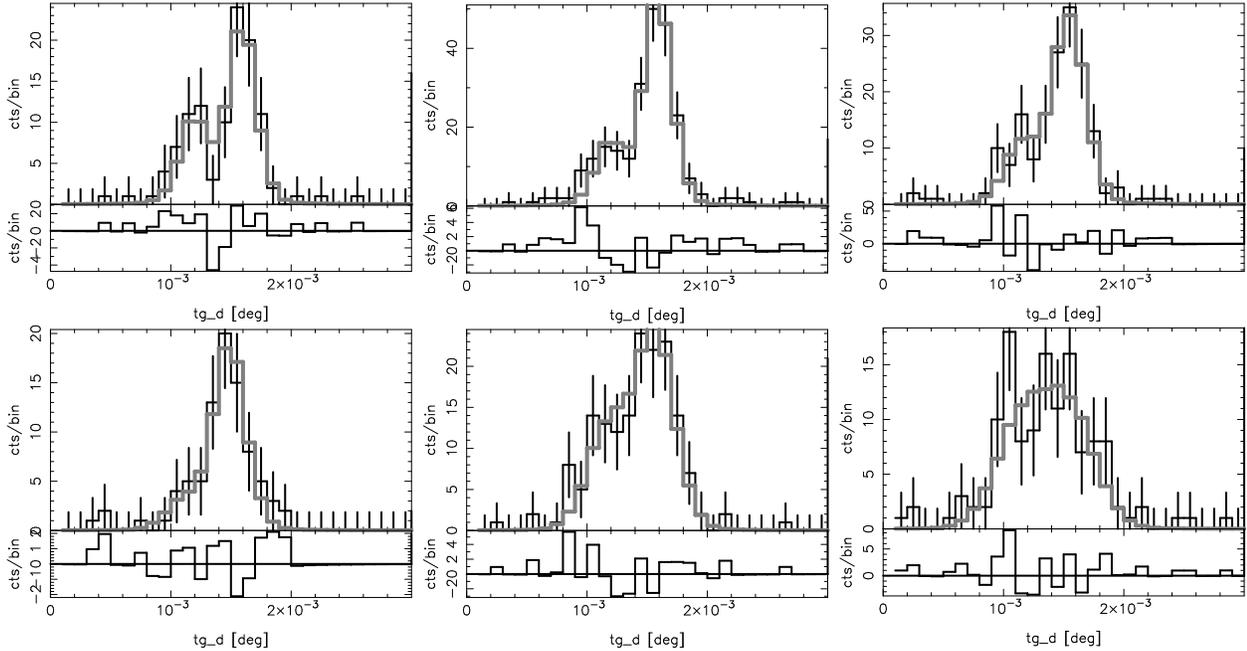

  \includegraphics[scale=0.3,angle=-90]{f8a.eps}
  \includegraphics[scale=0.3,angle=-90]{f8b.eps}
  \includegraphics[scale=0.3,angle=-90]{f8c.eps}\newline
  \includegraphics[scale=0.3,angle=-90]{f8d.eps}
  \includegraphics[scale=0.3,angle=-90]{f8e.eps}
  \includegraphics[scale=0.3,angle=-90]{f8f.eps}
  \caption{\small Cross-dispersion profiles
    of the MEG first-order spectrum of Hen~3-600 $AB$ (thin black
    histograms) in several spectral regions. A two-component model
    fit, based on the calibration model MEG cross-dispersion profile,
    is indicated in each panel by the thick gray histogram. Left to
    right, top row: \eli{Si}{14} 6 \AA;   \eli{Mg}{11}-\eli{Mg}{12}
    8--10\AA;  and the Fe and continuum region around 11 \AA.
    Left-to-right, bottom row: \eli{Ne}{10} 12 \AA;
    \eli{Fe}{17} 15--17 \AA; and \eli{O}{8} 20 \AA.
    The lower panel in each plot shows the residual $data-model$
    counts.   The zero-point of the x-axis is arbitrary; component $A$
    is on the left, $B$ is on the right.
    \label{fig:Xdisperse}}
\end{figure*}

The Hen~3-600 cross-dispersion profiles were fit with the sum of two
cross-dispersion calibration profiles, each corresponding to a single
source at the wavelength range of interest.  These wavelength ranges
were selected to be as narrow as feasible --- to extract information
on the relative strength of the two sources in individual spectral
lines where possible --- but wide enough to include counts sufficient
to allow a reliable fit. For each selected wavelength range, the
observed cross-dispersion profile generally shows two peaks which
correspond to the two main components of the Hen~3-600 binary.  The
separation between the two centroids was set equal to the known
separation between $A$ and $B$ ($1.4''$, or $3.9\times10^{-4}$
degrees) and not allowed to vary.  The Gaussian and Lorentzian width
parameters were also frozen at their known calibration values.  The
only free interesting parameters of the fit were the relative
normalizations of each component.  Figure~\ref{fig:Xdisperse} shows a
collection of the cross-dispersion fits for several line regions.

%
To derive the fluxes of individual spectral lines for each of the two
stars, we first determined the line fluxes in the combined ($A+B$),
dispersed spectrum. We then partitioned the combined line flux on the
basis of the weights derived from the best fit to the cross-dispersion
profiles in the corresponding wavelength region. The results are
listed in Table~\ref{tbl:abfluxes}. For individual bright lines such
as \eli{Ne}{10} $\lambda12.13$ and \eli{O}{8}
$\lambda18.97$, the cross-dispersion profile fitting 
can be performed over a narrow wavelength range, and the fluxes are
well determined for each binary component.  For weaker lines like
\eli{Si}{13} or the \eli{Ne}{9} intercombination and forbidden lines,
broad wavelength ranges were required for the profile fitting, and we
are only able to obtain the mean line and continuum contributions from
each component across a line complex. Other line fluxes that we
measured but which were used only in the emission measure
analysis (\S~\ref{sec:emd}) are listed in Table~\ref{tbl:lfluxemd}.

\subsubsection{Temperature sensitive line ratios of H-like and He-like ions}\label{sec:tsens}

In Section~\ref{sec:emd} we have used $EMD$ analysis to derive thermal
structure and element abundances for the composite spectrum of $A+B$,
using the fluxes of a large set of spectral lines.  Spectral
diagnostics based on a few strong spectral lines can be used as
independent checks on these $EMD$ results.  For instance, the He-like
to H-like Ly$\alpha$ line ratios can be used to derive an estimate of
the temperature of the emitting plasma. Even though the accurate
temperature determination from these ratios strictly holds only for
isothermal plasma, whereas the actual temperature distributions of
X-ray emitting plasma in stars are far from isothermal, these
diagnostics can provide useful comparisons between sources \citep[see,
e.g., ][]{Testa:al:2007b,Ness:gudel:al:2004}.

In Table~\ref{tbl:lineratios}, we summarize results for key line
ratios that are diagnostic of plasma temperature. Following the
cross-dispersion decomposition, a handful of these
temperature-sensitive line ratios are available separately for $A$ and
$B$. These ratios --- in particular, Ly$\alpha$/He-r line ratios of Ne
--- indicate that there were sharp plasma temperature differences
between $A$ and $B$ during our observations, with $B$ the hotter
component. The ratios of the H-like Ly$\alpha$ to He-like resonance
lines of Ne and O are indicative of plasma temperatures of $\sim2-3$
MK for component $A$ and $\sim7$ MK for component $B$. This is
consistent with the results from cross-dispersion profile analysis of
the He-like \eli{Ne}{9} and \eli{O}{7} triplet regions, which indicate
(albeit with a high degree of uncertainty) that $A$ is the fainter of
the two components in the former region but is comparable to component
$B$ in the latter, longer-wavelength spectral region.

\subsubsection{Elemental abundances}\label{sec:abund}

Using the cross-dispersion profile analysis, we can estimate the
abundances of Ne, Fe, Mg, and Si with respect to O for each stellar
component individually --- independently of (and in contrast to) the
$EMD$ analysis, in which the derived abundances
(Table~\ref{tbl:abundemd}) represent an average of the two components.
These diagnostics use the ratio of two combinations of lines of two
different elements whose combined emissivity curves have very similar
temperature dependence.  Therefore their ratio is largely independent
of the specific thermal distribution of the plasma and provides an
estimate of the ratio of the abundances of the relevant elements
\citep[see e.g.][]{Drake:Testa:2005,Garcia-Alvarez:al:2005}.

The cross-dispersion decomposition results for abundances are listed
in Table~\ref{tbl:abundratios}.  Consistent with the $EMD$ analysis,
we find both components of Hen~3-600 display anomalously high Ne/O
ratios and Fe deficiency, with component $A$ slightly less extreme in
terms of the latter. The Ne and Fe abundances and ratios to O of both
components are very similar to those of HD~98800
\citep{Kastner:Huenemoerder:al:2004} or BP~Tau
\citep{Robrade:Schmitt:2006}.  The Ne/O ratios, though, are not as
extreme as in the case of TW~Hya \citep{Kastner:02,
  Robrade:Schmitt:2006}.

Component $A$ displays evidence for enhanced abundances of Mg and Si
relative to O.  These anomalies, if confirmed, are similar to patterns
in TW~Hya, whose high-resolution X-ray spectrum \citep{Kastner:02}
reveals a relative overabundance of about 3 (compared to Solar ratios)
for both elements relative to oxygen.

\subsection{Density-sensitive line ratios of He-like ions}\label{sec:dsens}
%
\begin{figure}
  \epsscale{0.8}
  \centerline{\plotone{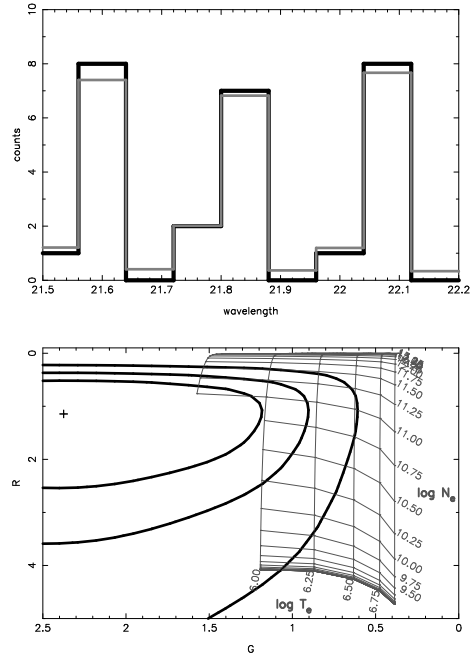}}
  \caption{\small Top: The \eli{O}{7} triplet counts spectrum (dark line) and
    best fit of Gaussians 
    folded through the instrument response (light-shaded line).  No
    background was subtracted and hence, uncertainties are Poisson.
    Bottom: confidence contours in the triplet density sensitive $R =
    f/i$ and temperature sensitive $G = (f+i)/r$ ratios.  Inner to
    outer contours are for 68\%, 90\%, and 99\% confidence levels.
    The grid \citep[after][]{Smith:01} gives the theoretical contours
    for constant densities and temperatures as labeled.
    \label{fig:otripconf}
  }
\end{figure}

\begin{figure}[t]
  \epsscale{0.8}
  \centerline{\plotone{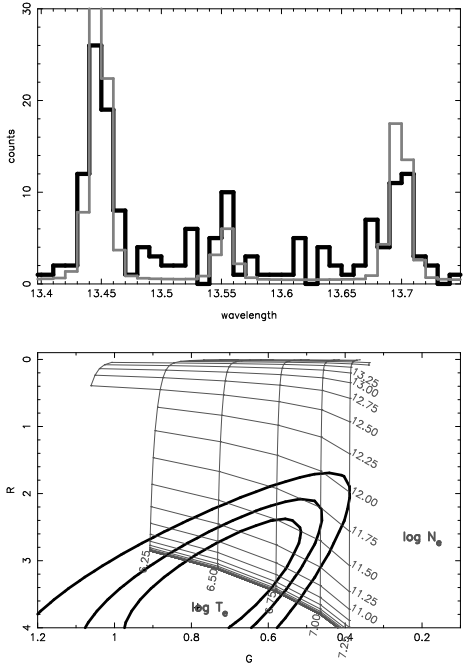}}
  \caption{\small Top: The \eli{Ne}{9} triplet spectrum and confidence
    contours. (See Figure~\ref{fig:otripconf} for details.)
    \label{fig:netrip}
  }
\end{figure}
%

Ratios of the forbidden ($f$) to intercombination ($i$) line
intensities within the triplet line complexes of He-like ions (e.g.,
\eli{Mg}{11}, \eli{Ne}{9}, \eli{O}{7}) are potential diagnostics of
plasma density.  For the combined ($A+B$) spectrum of Hen~3-600, the
$f/i$ ratio for the \eli{O}{7} triplet is significantly smaller than
the low-density limit of 3.9. The measured value, $f/i \sim 1\pm0.5$,
lies between the \eli{O}{7} $f/i$ line ratios displayed by HD~98800
($\sim3.5$) and TW~Hya ($\sim0.1$). It is not possible to infer plasma
densities independently and unambiguously for the two main binary
components of Hen~3-600, as the relatively poor photon counting
statistics limit the cross-dispersion profile fitting to broad
wavelength ranges that average over these triplet line complexes.
Table~\ref{tbl:abfluxes} and Figures~\ref{fig:zo}
and~\ref{fig:Xdisperse} suggest that components $A$ and $B$ contribute
roughly equally to the photon count rate in the \eli{O}{7} triplet
wavelength range, with $A$ perhaps the brighter component in this line
complex.

We computed confidence contours in the ratios $R = f/i$ and
$G=(f+i)/r$ (where $r$ is the intensity of the triplet resonance line)
directly from the Hen~3-600 $A+B$ spectrum
(Figure~\ref{fig:otripconf}). The results show that the 90\%
confidence limit extends to near the low density limit (middle
contour), but is still significantly above this limit.  However, the
$G$ ratio is unrealistically large, extending to temperatures outside
the theoretical range of sensitivity, indicating some peculiarity in
the emission, or a problem with the atomic data.  If we consider only
the $f/i$ ratio ($R$), then $\log n_e [\mathrm{cm\mthree}] \sim
10.75$, with the 90\% confidence interval at $\log T = 6.0$ ranging
from 10.25 to 11.25.  The \eli{Ne}{9} $f/i$ line ratio indicates an
upper limit of $\log n_e [\mathrm{cm}\mthree] \le 11.75$ (90\%
confidence), consistent with \eli{O}{7} (Figure~\ref{fig:netrip}).
The \eli{Ne}{9} triplet region has Fe line blends, but since Fe is
relatively weak in the spectrum of Hen~3-600, the $R$ ratio for
\eli{Ne}{9} is essentially uncontaminated.  The \eli{Mg}{11} triplet
is compromised by weak signal, \eli{Ne}{10} H-Ly$\alpha$ series blends
which bracket the intercombination line, and significant continuum.

\subsection{\eli{O}{7} ratios:  
  $N_H$, abundances, and binarity}\label{sec:ovii} 

In principle, given the theoretical ratio of flux in \eli{O}{7}
He-Ly$\beta$ (18.63 \AA) to the flux in the triplet complex (22 \AA),
we should be able to determine the line-of-sight absorbing column of
neutral hydrogen $N_H$.  As $N_H$ increases, the ratio of 18.63 \AA\ 
to 22 \AA\ apparent flux should increase above the theoretical value
of 0.06, due to preferential absorption of longer-wavelength photons.
Indeed, the ratio we observe for He-Ly$\beta$ to the sum of the
triplet is about 2.5-3 times the theoretical value of 0.06.  The
measurement uncertainty is large, however; the 90\% confidence
interval reaches into the range 0.04-0.08 in the ratio defined by its
temperature dependence over the peak emissivity of \eli{O}{7}
(approximately $\log T = 6.2 $ to 6.5).

The binarity of the system introduces complications, however, since
the dispersed spectra of components $A$ and $B$ are blended.  There is
no reason {\it a priori} that any local $N_H$ be the same for the two
components; indeed, we would expect $N_H$ toward $A$ to be larger than
that toward $B$, given that only the former component has a detectable
mass of circumstellar dust.  In addition, the line fluxes are
determined by the product of the emission measure and the oxygen
abundance, either of which could also differ between the two stars.

One empirical constraint, from the cross-dispersion profile fitting
(\S \ref{sec:xdisp}), is that the emergent flux near 20 \AA\ is about
equal for $A$ and $B$.  Assuming that this reflects similar flux in
the \eli{O}{7} transitions from each star, the ratio of \eli{O}{7}
fluxes to the ratio of theoretical emissivities is then only a
function of $N_H$ for each stellar component.  Contours of constant
ratios vs.\ $N_H(A)$ and $N_H(B)$ (not shown) indicate that $N_H$ must
be $> 2\times10^{21}$ for either component to drive the observed ratio
to values $>2$.  However, such a large value of $N_H$ is inconsistent
with the relatively strong \eli{N}{7} line, as discussed earlier
(\ref{sec:emd}).  We conclude that the deviation of the observed ratio
of 18.63 \AA\ to 22 \AA\ flux from the theoretical value is most
likely due to poor photon counting statistics. 

%
\section{Discussion}\label{sec:disc}

\subsection{He-like triplet line ratios}

\begin{figure}[t]
  \includegraphics[scale=0.45,angle=90]{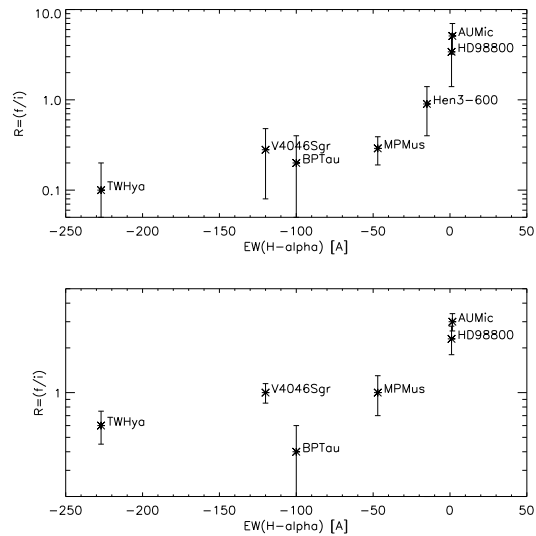}
  \caption{\small Equivalent width ($EW$) of H$\alpha$ vs.\ the
    forbidden to intercombination line ratio $f/i$ within
    triplet lines of the He-like ions \eli{O}{7} (top
    panel) and \eli{Ne}{9} (bottom panel). The values of
    H$\alpha$ $EW$ were obtained from \citet{Webb:Zuckerman:al:1999}
    (Hen 3-600, TW Hya, HD 98800),
    \citet{GregorioHetem:al:1992} (MP Mus),
    \citet{Byrne:1986} (V4046 Sgr), \citet[][and
    references therein]{Schmitt:Robrade:al:2005} (BP Tau), and
    \citet{BarradoyNavascues:al:1999} (AU Mic).}
  \label{fig:R_vs_Halpha}
\end{figure}

\begin{figure}[ht]
  \includegraphics[scale=0.45,angle=90]{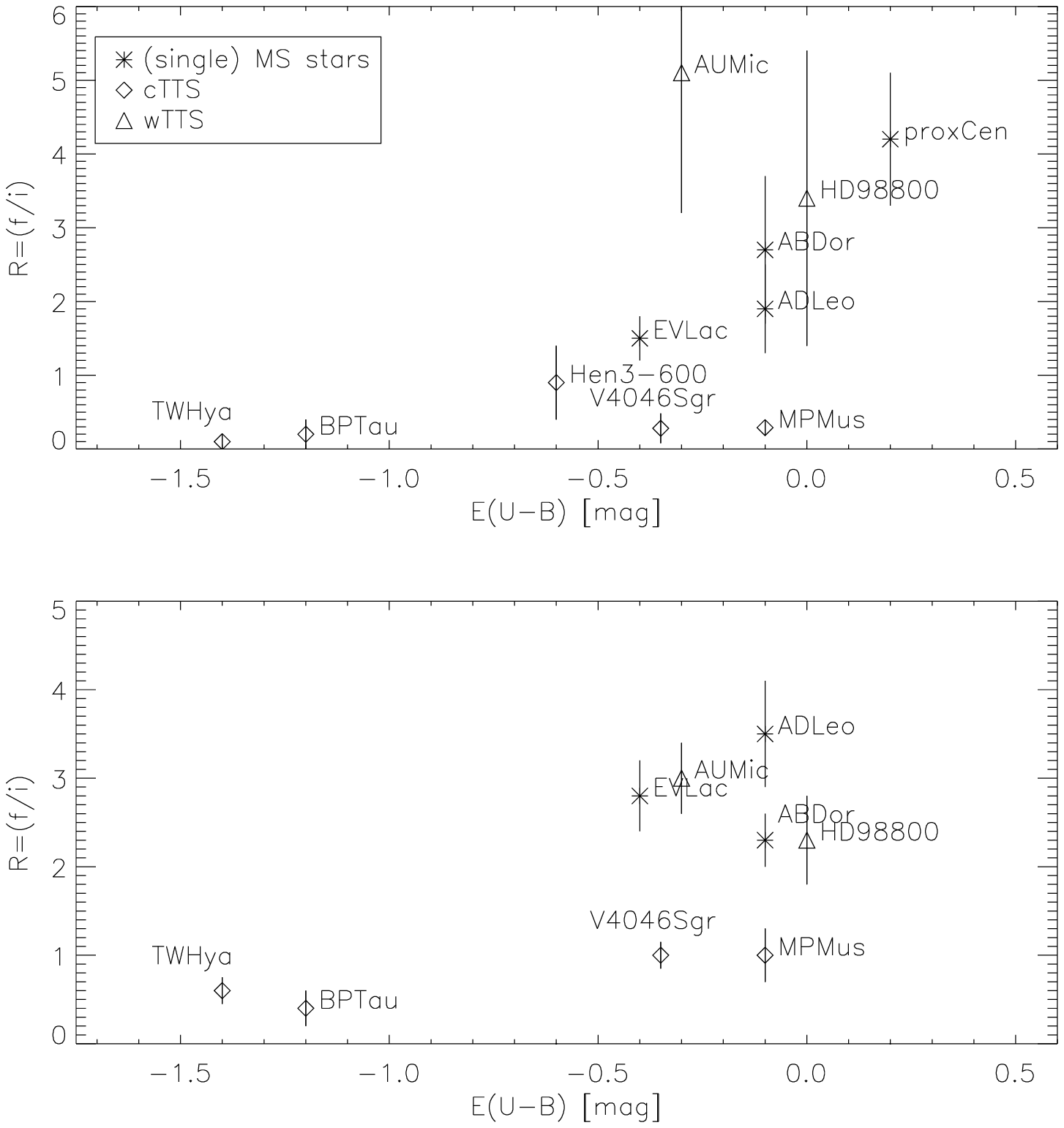}
  \caption{\small Ultraviolet excess $E(U-B)$ vs.\ $f/i$ within triplet lines
    of the He-like ions \eli{O}{7} (top panel) and \eli{Ne}{9}
    (bottom panel) as measured for chromospherically active, single
    main sequence stars (asterisks) and low-mass, pre-main sequence
    stars (diamonds, cTTS; triangles, wTTS). The values of $E(U-B)$
    were estimated from $U$ and $V$ photometry listed in
    \citet{delaReza:al:1989} (Hen~3-600, TW~Hya),
    \citet{Hutchinson:Evans:al:1990} (V4046 Sgr),
    \citet{GregorioHetem:al:1992} (HD 98800, MP Mus), and SIMBAD (all
    other stars) by taking into account intrinsic $(U-B)$ colors
    appropriate for each star's spectral type \citep{Johnson:1966}.}
  \label{fig:R_vs_EUB}
\end{figure}

Both TW~Hya and Hen~3-600 exhibit H$\alpha$ emission line equivalent
widths and UV excesses (Table~\ref{tbl:stars}) that are larger than
typically observed for chromospherically active stars.  Such large
H$\alpha$ emission intensities and UV excesses are generally
interpreted as evidence for accretion.  In
Figures~\ref{fig:R_vs_Halpha} and \ref{fig:R_vs_EUB}, respectively, we
plot the equivalent widths of H$\alpha$ and ultraviolet excesses
$E(U-B)$ vs.\ $f/i$ ratios of triplet lines of He-like ions for
pre-main sequence stars and (in Figure~\ref{fig:R_vs_EUB}) for
chromospherically active main sequence stars.  These figures make
clear that:
\begin{enumerate}

\item for T Tauri stars, H$\alpha$ equivalent widths and X-ray $f/i$
  line ratios are well correlated;

\item the cTTS systems (TW Hya, BP Tau, V4046 Sgr, and MP Mus) have
  anomalously small values of $f/i$;

\item the properties of the apparently nonaccreting (wTTS) systems AU
  Mic and HD 98800 more closely resemble those of coronally active
  main sequence stars than actively accreting pre-main sequence stars
  \citep[as also noted by][]{Kastner:Huenemoerder:al:2004};

\item the weakly accreting Hen~3-600 appears as an intermediate case,
  lying between the cTTS on the one hand, and wTTS and coronally
  active main sequence stars on the other.

\end{enumerate}
%

\clearpage

The presence of intense UV fields generated in accretion shocks that
lie in close proximity to the X-ray-emitting regions of cTTS may
affect the $f/i$ ratios of such systems, via radiative pumping of ions
out of the metastable state that leads to forbidden line emission.  If
such radiative pumping is important, clearly this would compromise the
utility of the $f/i$ ratio as a density diagnostic.  However, models
suggest that that such UV pumping contributes negligibly to the
observed line ratios of He-like ions for TW~Hya and Hen~3-600.
Specifically, to produce the $f/i$ ratios observed for Hen~3-600
(\eli{O}{7}) and TW~Hya (\eli{Ne}{9}), the UV fields would have to be
characterized by temperatures of $\sim10^4$ K and $\sim1.5\times10^4$
K, respectively \citep{Porquet:01}.  In contrast, the extant UV
continuum data for these stars indicates temperatures well below
$\sim10^4$ K in both cases
\citep{Costa:Lago:al:2000,Muzerolle:Calvet:2000}.  Furthermore, there
is no apparent correlation between $f/i$ and ultraviolet excess, among
cTTS systems (Figure~\ref{fig:R_vs_EUB}).  Hence it appears the plasma
densities characterizing cTTS systems and Hen~3-600 are anomalously
high among X-ray-emitting, late-type stars observed thus far at high
spectral resolution.  Recent results reported by
\citet{Ness:Schmitt:2005} concerning anomalous Fe line ratios in the
X-ray spectrum of TW~Hya support this assertion.

\subsection{X-ray temperatures and origins in $A$ and $B$}

Strong low-$T$ ($\log T\sim 6.5$) emission is apparent in the \hetgs\ 
spectrum of Hen~3-600.  This result is obtained both from the \emd\ 
modeling results for the combined ($A+B$) system and from temperature
sensitive Ne and O line ratios for component $A$ as derived from
cross-dispersion decomposition (Table~\ref{tbl:lineratios}).  The
similarly low plasma temperatures measured for Hen~3-600-$A$ and TW~Hya
point to the possibility that, in both cases, the X-ray emission is
derived (at least in part) from shocks.  In this regard, we note that
the X-ray flux from Hen~3-600-$A$ was relatively constant during our
observation, consistent with a shock origin.  Furthermore, the
temperature derived from \eli{O}{7} line ratios in the spectrum of $A$
($\sim3$ MK) is similar to that measured for other accreting TTS,
whereas the \eli{Ne}{9} and \eli{O}{7} line ratio temperatures of $B$
($\sim5-7$ MK) more closely resembles those of non-accreting TTS
\citep{Telleschi:Gudel:Briggs:Audard:Scelsi:2007}.  

There is additional evidence that the X-ray emission from component
$B$ is dominated by magnetically heated (as opposed to shock-heated)
plasma.  Specifically, it displayed a hotter spectrum than $A$ as
determined both in the continuum and in line ratios, and it displayed
a large flare.  The flare was somewhat peculiar, however, in that the
spectral hardness showed no strong change as is typical for coronally
active stars.  This circumstantial evidence may simply be due to
statistics; further observations may detect a flare on $A$.  In the
Taurus molecular cloud, cTTS and wTTS do show similar variability
\citep{Stelzer:Flaccomio:al:2007}.  This implies that even strongly
accreting sources still probably have a significant emission component
from magnetically heated plasma.  As any accretion-generated emission
component weakens, magnetic coronal activity will become more
apparent.  In this middle regime, high-resolution (line-based)
diagnostics of temperatures and abundances are needed.

\subsection{Abundances}

Whereas TW~Hya displays evidence for metal (O, Fe, Mg, Si) depletion
that has been attributed to grain growth in its circumstellar disk
\citep{Stelzer:Schmitt:2004, Drake:al:2005}, the abundance patterns
found in the X-ray spectrum of Hen~3-600 are ambiguous in this regard:
Fe is depleted, but the absolute abundance ratio Ne/O is $\sim0.4$
($\sim2.8$ relative to the solar Ne/O ratio), very much in line with
the mean Ne/O ratio of coronally active stars
\citep{Drake:Testa:2005}.

If, however, we consider only the results from high-resolution X-ray
spectroscopy, it is not at all clear that there is such strong
evidence for depletion.  The Si and Mg abundances in TW~Hya are about
Solar according to \cite{Kastner:02}, and while those authors stated
problems regarding the \eli{Mg}{11} model, \eli{Si}{14}, \eli{Si}{13},
and \eli{Mg}{12} are well described by their adopted emission measure
and abundances.  Thus, the largest difference between Hen 3-600 and
TW~Hya is in their Ne/O ratios which are about 3 for Hen~3-600 (as
well as for HD~98800, BP~Tau, and CR~Cha), and about half that of
TW~Hya (using the \citet{Anders:89} scale).  The abundances of Mg and
Si themselves appear to be substantially lower in Hen~3-600 than in
TW~Hya.

The pattern for Hen~3-600-$AB$ is similar to that found in other
active stars from high resolution spectra, but which are also seen in
the much younger Orion Nebula Cluster stars from analysis of
low-resolution CCD spectra by \citet{Maggio:Flaccomio:al:2007}.  The
large sample of Orion stars allowed those authors to apply independent
statistical tests.  Our values for Hen~3-600-$AB$ fall within the
$1\sigma$ boxes shown in their Figure~12, which in turn overlap values
for other stars.

Considering the decomposition by component star, the Si and Mg
abundances relative to O in component $A$ are, if anything, enhanced
with respect to Solar ratios, and are more like TW~Hya than is
component $B$.  Indeed, in light of the fact that the two most rapidly
accreting stars in the TWA, TW~Hya and Hen~3-600, both display
evidence for grain processing \citep{Uchida:Calvet:al:2004}, that only
TW~Hya displays an anomalous Ne/O ratio, and that both TW~Hya and
Hen~3-600-$A$ have high Si/O and Mg/O values, it appears that
circumstellar dust grain evolution may play an important role in
determining X-ray abundance patterns, but these patterns do not seem
to be what was previously expected for Si and Mg.  It is clear that
better high-resolution spectra are required to provide much better
statistics and models for the important Mg and Si features.  Our
uncertainties on the $A$-$B$ decomposition are quite large due
primarily to the overall signal level, but are not limited by the
spatial separation of the two stars.

\section{Conclusions}

The results presented here establish that Hen~3-600, like TW~Hya,
stands out from classical coronal sources, in terms of both its
\eli{O}{7} triplet spectrum and its soft X-ray (low temperature
emission measure) excess.  Specifically, among the several dozen stars
with \eli{O}{7} triplet line ratios thus far measured by \chan/\hetgs\
or \xmm/\rgs, Hen~3-600, TW~Hya, BP Tau, V4046~Sgr, and MP~Mus -- all
cTTS systems --- are the only stars with $R=f/i$ less than $\sim1.5$
\citep{Kastner:02,Testa:Drake:al:2004b, Ness:Schmitt:2005,
  Robrade:Schmitt:2006, Gunther:Liefke:al:2006,
  Argiroffi:Maggio:Peres:2007}. Meanwhile, soft X-ray excesses have
been detected in the spectra of a number of accreting TTS systems and
are not observed in non-accreting systems
\citep{Telleschi:Gudel:Briggs:Audard:Scelsi:2007}. While the
\eli{O}{7} $f/i$ ratio and the $EMD$ peak at $\log{T_X} \sim 6.5$
measured for Hen~3-600 are not as extreme as those found for TW~Hya
\citep{Kastner:02}, it is notable that these two objects --- the only
TWA stars known to be actively accreting --- share these unusual X-ray
spectral properties.

Hence, it seems that the soft X-ray emission from Hen~3-600-$A$, like
that from TW Hya and few other cTTS observed to date at high spectral
resolution in X-rays, may be directly attributable to accretion shocks
\citep{Kastner:02, Stelzer:Schmitt:2004, Gunther:Liefke:al:2006}.
Indeed --- while conclusions concerning Hen~3-600 are somewhat
tentative at present, due to the poor photon counting statistics in
our \hetgs\ spectroscopy --- the X-ray-derived plasma densities of
TW~Hya and Hen~3-600 ($\log{n_e} \approx 12.5$ and $\log{n_e} \approx
11$, respectively) appear to scale with their accretion rates as
derived from H$\alpha$ line profiles and UV continuum fluxes:
$\sim5\times10^{-10} M_\odot$ yr$^{-1}$ and $\sim10^{-11} M_\odot$
yr$^{-1}$, respectively \citep{Muzerolle:Calvet:2000}.  Alternatively,
this low-$T$ emission perhaps may be the result of shocks in
collimated, disk-driven outflows from these stars.  Such soft,
constant X-ray emission components have recently been detected in
\chan\ and \xmm\ CCD spectroscopy of a handful of cTTS disk/jet
systems, such as the Beehive proplyd in Orion
\citep{Kastner:Franz:al:2005} and DG~Tau
\citep{Gudel:Telleschi:al:2007,Gudel:Skinner:al:2005}.

All of the emission signatures of Hen~3-600-$A$ that are potential
diagnostics of accretion activity --- from its X-ray line ratios,
through its UV excess and H$\alpha$ emission line strengths, to its
weak near-infrared excess --- are fully consistent with its status as
a transition object, placed roughly midway between rapidly accreting,
classical T Tauri stars and non-accreting, weak-lined T Tauri stars.
These results speak to the need for additional high-resolution X-ray
spectroscopy of T Tauri stars spanning a wide range of accretion
states.

\acknowledgements
Support for this research was provided by NASA/CXO
  grant GO4--5012X to RIT, and by NASA through the Smithsonian
  Astrophysical Observatory (SAO) contract SV3-73016 to MIT for the
  \chan\ X-Ray Center and Science Instruments.  We thank Dr.~K.
  Ishibashi for assistance with the \hetgs\ LSF calibration data.
  This research has made use of the SIMBAD database, operated at CDS,
  Strasbourg, France.

{\it Facilities:} \facility{CXO(HETG)}.

%

\input{ms.bbl}
  \begin{deluxetable}{ccccc}
    \tabletypesize{\footnotesize}
    \tablewidth{0pt}
    \tablecaption{Properties of Hen~3-600, HD~98800, and TW~Hya}
    \tablehead{
      \colhead{}& 
      \colhead{}& 
      \colhead{EW(H$\alpha$)}& 
      \colhead{$U-B$}& 
      \colhead{$E(U-B)$}\\
      \colhead{Object}& 
      \colhead{Sp.\ Type}& 
      \colhead{(\AA)}& 
      \colhead{(mag)}& 
      \colhead{(mag)}
    }
    \startdata
    HD~98800   & K5\tablenotemark{a}    & 0              & 1.1     & 0.0 \\
    Hen~3-600  & M3\tablenotemark{b},M3.5 & $-22,-7$     & 0.7     & $-0.5$ \\
    TW~Hya     & K7        & $-220$         & $-0.3$  & $-1.4$ \\
    \enddata
    \tablenotetext{a}{Hierarchical quadruple system.} 
    \tablenotetext{b}{Spectroscopic binary.}
    \label{tbl:stars}
  \end{deluxetable}

\input{tab2}

\clearpage
\input{tab3}

\clearpage
\input{tab5}

\input{tab4}

\input{tab6}

\end{document}

%% file: tab2.tex
\begin{deluxetable}{lcrrr}
\tablecolumns{5}
\tabletypesize{\small}
\tablecaption{Line Fluxes Measured from $(A+B)$ \label{tbl:lfluxemd} }
\tablewidth{0pt}
\tablehead{
  \colhead{Ion}&
  \colhead{$\log T_\mathrm{max}$}&
  \colhead{$\lambda$}&
  \colhead{$f$\tablenotemark{a}}&
  \colhead{$f_\mathrm{pred}$\tablenotemark{b} } \\
  \colhead{}&
  \colhead{[K]}&
  \colhead{[\AA]}&
  \multicolumn{2}{c}{ $10^6\,[\mathrm{ph\,cm\mtwo\,s\mone}]$ }
}
\startdata 
\eli{Fe}{25}&  7.8&   1.850&  0.80 (0.60)& 0.28\\
\eli{Ar}{18}&  7.7&   3.749&  0.36 (0.32)& 0.17\\
\eli{Ar}{17}&  7.4&   3.935&  0.39 (0.32)& 0.31\\
\eli{S}{15}&  7.3&   4.075&  0.23 (0.30)& 0.05\\
\eli{S}{16}&  7.6&   4.739&  0.70 (0.60)& 0.77\\
\eli{S}{15}&  7.2&   5.026&  1.94 (0.60)& 1.19\\
\eli{S}{15}&  7.2&   5.060&  0.60 (0.50)& 0.28\\
\eli{S}{15}&  7.2&   5.101&  0.90 (0.50)& 0.41\\
\eli{Si}{14}&  7.4&   5.242&  0.00 (0.10)& (0.19)\\
\eli{Si}{14}&  7.4&   6.182&  1.93 (0.50)& 1.38\\
\eli{Si}{13}&  7.0&   6.648&  2.34 (0.80)& 1.77\\
\eli{Si}{13}&  7.0&   6.687&  1.60 (0.70)& (0.34)\\
\eli{Si}{13}&  7.0&   6.740&  1.30 (0.57)& 0.73\\
\eli{Mg}{12}&  7.2&   7.109&  0.32 (0.22)& 0.23\\
\eli{Mg}{11}&  6.9&   7.305&  0.06 (0.13)& (0.03)\\
\eli{Mg}{12}&  7.2&   8.421&  2.18 (0.60)& 1.74\\
\eli{Mg}{11}&  6.8&   9.169&  2.20 (0.70)& 1.45\\
\eli{Mg}{11}&  6.8&   9.314&  1.33 (0.50)& (0.69)\\
\eli{Ne}{10}&  7.0&  10.239&  3.50 (0.90)& 3.49\\
\eli{Fe}{23}&  7.2&  10.979&  1.19 (0.58)& 0.61\\
\eli{Fe}{24}&  7.4&  11.177&  2.36 (0.70)& 0.81\\
\eli{Fe}{18}&  6.8&  11.533&  0.28 (0.97)& (0.41)\\
\eli{Ne}{9}&  6.6&  11.543&  1.60 (0.96)& 2.09\\
\eli{Fe}{23}&  7.2&  11.741&  1.84 (0.74)& 1.32\\
\eli{Fe}{22}&  7.1&  11.774&  1.38 (0.67)& 1.45\\
\eli{Ne}{10}&  6.9&  12.134& 33.35 (3.49)& 26.31\\
\eli{Fe}{17}&  6.7&  12.269&  1.20 (0.90)& 1.35\\
\eli{Fe}{21}&  7.1&  12.281&  5.08 (1.30)& 3.25\\
\eli{Ne}{9}&  6.6&  13.447& 22.55 (3.19)& 17.51\\
\eli{Ne}{9}&  6.6&  13.553&  3.69 (2.99)& (2.84)\\
\eli{Ne}{9}&  6.6&  13.699& 10.58 (2.99)& (9.03)\\
\eli{Fe}{18}&  6.8&  14.205&  5.75 (1.90)& 4.48\\
\eli{Fe}{18}&  6.8&  14.377&  4.31 (1.92)& 1.15\\
\eli{O}{8}&  6.7&  14.809&  2.31 (1.79)& 1.22\\
\eli{Fe}{17}&  6.7&  15.014& 15.85 (4.98)& 14.38\\
\eli{Fe}{17}&  6.7&  15.261&  5.98 (2.99)& 4.06\\
\eli{O}{8}&  6.7&  16.006& 10.96 (4.48)& 8.89\\
\eli{Fe}{18}&  6.8&  16.071&  5.18 (4.18)& 1.63\\
\eli{Fe}{17}&  6.7&  16.780&  9.36 (3.98)& 6.52\\
\eli{Fe}{17}&  6.7&  17.051& 14.44 (5.68)& 7.73\\
\eli{Fe}{17}&  6.7&  17.096&  8.37 (5.88)& 7.25\\
\eli{O}{7}&  6.4&  18.627& 18.25 (5.09)& 3.16\\
\eli{O}{8}&  6.7&  18.969& 82.75 (11.96)& 68.28\\
\eli{O}{7}&  6.3&  21.605& 25.50 (9.56)& 25.14\\
\eli{O}{7}&  6.3&  21.805& 31.18 (11.46)& (3.41)\\
\eli{O}{7}&  6.3&  22.092& 34.66 (11.85)& (14.03)\\
\eli{N}{7}&  6.5&  24.780& 33.87 (13.03)& 25.07\\
\enddata

\tablenotetext{a}{\ Measured line fluxes for $\heg+\meg$ first orders,
  with $1\sigma$ uncertainties.  Fluxes are for $N_H = 0.0$.}

\tablenotetext{b}{\ Line fluxes predicted by the emission measure and
  abundance model.  Lines which are known to be density sensitive
  (He-like $i$ and $f$ lines), or which had very large residuals, were
  not used in the emission measure fit and are enclosed in
  parentheses.}

\end{deluxetable}

%% file: tab3.tex
\begin{deluxetable}{lcrrrrr}
\tablecolumns{6} 
\tabletypesize{\small}
\tablecaption{Decomposition of fluxes from cross dispersion profiles.
  \label{tbl:abfluxes}}
\tablewidth{0pt}
\tablehead{ 
  \colhead{Ion}&
  \colhead{$\log T_\mathrm{peak}$}&
  \colhead{$\lambda$}&
  \colhead{$(A/B)$\tablenotemark{a}}&
  \colhead{$A$} &
  \colhead{$B$}\\ \cline{5-6}
  \colhead{} &
  \colhead{[K]} &
  \colhead{[\AA]} &
  \colhead{} & 
  \multicolumn{2}{c}{flux ( $10^{6}\mathrm{[ph\,cm\mtwo\,s\mone]}$) }
}
\startdata 
Fe\,{\sc xxv}\tablenotemark{b}   & 7.8 & 1.854  & \hfill\ldots\hfill\ 
& \hfill\ldots\hfill\   & 0.8 (0.6)\tablenotemark{c} \\
Si\,{\sc xiv}   & 7.2 & 6.182  & 0.55 (0.15) &    0.68 (0.18) & 1.25 (0.32) \\
Si\,{\sc xiii}  & 7.0 & 6.648  & 0.55 (0.15) &    0.83 (0.27) & 1.51 (0.48) \\
Si\,{\sc xiii}  & 7.0 & 6.687  & 0.55 (0.15) &    0.57 (0.25) & 1.03 (0.45) \\
Si\,{\sc xiii}  & 7.0 & 6.740  & 0.55 (0.15) &    0.46 (0.20) & 0.84 (0.38) \\
Mg\,{\sc xii}   & 7.0 & 8.421  & 0.30 (0.13) &    0.50 (0.14) & 1.68 (0.45) \\
Mg\,{\sc xi}    & 6.8 & 9.169  & 0.30 (0.13) &    0.51 (0.18) & 1.69 (0.55) \\
Mg\,{\sc xi}    & 6.8 & 9.314  & 0.30 (0.13) &    0.31 (0.12) & 1.02 (0.36) \\
Ne\,{\sc x}     & 6.8 & 10.239 & 0.30 (0.14) &    0.73 (0.21) & 2.76 (0.70) \\
Ne\,{\sc x}     & 6.8 & 12.134 & 0.15 (0.13) &    4.40 (0.40) & 29.0 (3.0) \\
Ne\,{\sc ix}    & 6.6 & 13.447 & 1.0  (0.9)  &    11.3 (2.0) & 11.3 (2.0) \\
Ne\,{\sc ix}    & 6.6 & 13.553 & 0.66 (0.27) &    1.5  (1.3) & 2.2  (1.9) \\
Ne\,{\sc ix}    & 6.6 & 13.699 & 0.66 (0.27) &    4.2  (1.3) & 6.4  (1.9) \\
Fe\,{\sc xvii}  & 6.7 & 15.014 & 0.50 (0.19) &    5.3  (1.7) & 10.6 (3.5) \\
Fe\,{\sc xvii}  & 6.7 & 15.261 & 0.50 (0.19) &    2.0  (0.9) & 4.0  (2.0) \\
O\,{\sc viii}   & 6.5 & 16.006 & 0.50 (0.19) &    3.7  (1.5) & 7.3  (2.8) \\
Fe\,{\sc xviii} & 6.8 & 16.071 & 0.50 (0.19) &    1.7  (1.2) & 3.5  (2.3) \\
Fe\,{\sc xvii}  & 6.7 & 16.780 & 0.50 (0.19) &    3.1  (1.5) & 6.3  (2.5) \\
Fe\,{\sc xvii}  & 6.7 & 17.051 & 0.50 (0.19) &    4.8  (1.9) & 9.7  (3.7) \\
Fe\,{\sc xvii}  & 6.7 & 17.096 & 0.50 (0.19) &    2.8  (1.9) & 5.6  (3.8) \\
O\,{\sc viii}   & 6.5 & 18.969 & 0.75 (0.40) &    34.0 (6.0) & 49.0 (7.0) \\
O\,{\sc vii}    & 6.3 & 21.602 & 2.0  (1.5)  &    13   (7) & 6.5  (3) \\
O\,{\sc vii}    & 6.3 & 21.804 & 2.0  (1.5)  &    21   (7) & 11   (4) \\
O\,{\sc vii}    & 6.3 & 22.098 & 2.0  (1.5)  &    19   (8) & 9.7  (4) \\
\enddata
\tablenotetext{a}{$A$ and $B$ refer to the photon fluxes of the
  stellar components.}
\tablenotetext{b}{Flux measured from the 0th order spectrum of source B; the feature is
  not detected in the 0th order spectrum of source A.}
\tablenotetext{c}{Quantities in  parentheses are the $3\sigma$ confidence intervals.}
\end{deluxetable}

%% file: tab5.tex
%
\begin{table}[t]
\begin{center}
\caption{Abundances from emission measure fit to
  $A+B$} \label{tbl:abundemd}
\begin{tabular}{lrr}
\tableline
Element&
$A/A_\odot$\tablenotemark{a}&
$A/A_\mathrm{O}$\tablenotemark{b}\\
\tableline
N&   1.62 (0.74)\tablenotemark{c}& 3.86 (1.90)\\
O&   0.42 (0.08)& 1.00 \hfill\-\ldots\hfill\ \\
Ne&  1.17 (0.10)& 2.79 (0.59)\\
Mg&  0.29 (0.07)& 0.68 (0.22)\\
Si&  0.39 (0.10)& 0.92 (0.29)\\
S&   0.75 (0.31)& 1.78 (0.81)\\
Fe&  0.20 (0.03)& 0.48 (0.12)\\
\tableline
\end{tabular}
\end{center}
\tablenotetext{a}{Abundance relative to Solar values of \citet{Anders:89}.}
\tablenotetext{b}{Abundance ratio relative to Solar oxygen abundance
  ratio, or $(A/A_\odot) / (A_\mathrm{O}/A_\mathrm{O_\odot})$.}
\tablenotetext{c}{Values in parentheses are $1\sigma$ uncertainties.}
\end{table}

%% file: tab4.tex
\begin{table}[t]
\begin{center}
\caption{Line ratios (photon units).}\label{tbl:lineratios}
\begin{tabular}{ccrrrr}
\tableline
Element &
Feature &
ratio($A$)  &
ratio($B$)&
$T(A)$&
$T(B)$  \\
&
&
($1\sigma$)&
($1\sigma$)&
\multicolumn{2}{c}{[MK] (90\% confidence)}\\
\tableline
Si&   Ly$\alpha$/He-$r$      & 0.82 (0.34)  &  $[A=B]$& 7.8 (6.2--8.2)& $[A=B]$\\
Mg&   Ly$\alpha$/He-$r$      & 0.99 (0.42)  &  $[A=B]$& 8.0 (6.2--9.0)& $[A=B]$\\
Ne&   Ly$\alpha$/He-$r$      & 0.39 (0.08)  & 2.56 (0.50)& 3.8 (3.4--4.0)& 6.7 (6.0--7.0)\\
 O&   Ly$\alpha$/He-$r$      & 2.6  (1.5)   & 7.5 (3.6)  & 3.2 (2.6--3.8)& 5.0 (3.8--5.8) \\
\tableline
\end{tabular}
\end{center}
\end{table}

%% file: tab6.tex

\begin{table}[t]
\begin{center}
  \caption{Abundance ratios (relative to Solar ratios) from
    temperature-insensitive line ratios.\label{tbl:abundratios}}
\begin{tabular}{lrrr}
\tableline
Ratio& $A$\tablenotemark{a}  & $B$ & $A+B$ \\
\tableline
Ne/O&  3.68 (0.89)\tablenotemark{b}&   3.43 (0.67)&   3.56 (0.56)\\
Mg/O&  2.02 (0.82)&   1.13 (0.40)&   1.26 (0.46)\\
Si/O&  4.79 (2.65)&   1.47 (0.93)&   2.02 (1.40)\\
Fe/O&  0.69 (0.26)&   0.31 (0.12)&   0.39 (0.14)\\
\tableline
\end{tabular}
\end{center}
\tablenotetext{a}{Ratio of abundances relative to the Solar ratio from
\citet{Anders:89}.}
\tablenotetext{b}{Values in parenthesis are $1\sigma$ uncertainties.}
\end{table}